 \documentstyle[12pt]{article} 
\begin{document} 
 \begin{center} 
 \vspace*{\fill} 

 {\LARGE Classical Signature Change in the \\[3mm] 
 Black Hole Topology} 
 \vspace*{1cm}\\ 

 {\bf Charles Hellaby, Ariel Sumeruk and G.F.R. Ellis} 

 {\small 
 Department of Applied Mathematics, \\ 
 University of Cape Town, \\ 
 Rondebosch, \\ 
 7700, \\ 
 South Africa} \\ 
 E-mail: {\tt cwh@appmath.uct.ac.za} \\ 
 \vfill 

 {\it Appeared in:~~~ Int. J. Mod. Phys. D, {\bf 6}, 211-38 (1997)} \\
 {\it gr-qc/9907042}
 \vfill

 {\bf \large Abstract} \\[6mm] 

 \parbox{12cm}{ 
 Investigations of classical signature change have generally envisaged 
applications to cosmological models, usually a 
 Friedmann-Lema\^{\i}tre-Robertson-Walker model.  The purpose has been 
to avoid the inevitable singularity of models with purely Lorentzian 
signature, replacing the neighbourhood of the big bang with an 
initial, singularity free region of Euclidean signture, and a 
signature change.  We here show that signature change can also avoid 
the singularity of gravitational collapse.  We investigate the process 
of
 re-birth of Schwarzschild type black holes, modelling it as a double 
signature change, joining two universes of Lorentzian signature 
through a Euclidean region which provides a `bounce'.  We show that 
this process is viable both with and without matter present, but 
realistic models 
 --- which have the signature change surfaces hidden inside the 
horizons 
 --- require
 non-zero density.  In fact the most realistic models are those that 
start as a finite cloud of collapsing matter, surrounded by vacuum.  
We consider how geodesics may be matched across a signature change 
surface, and conclude that the particle `masses' must jump in value. 
 This scenario may be relevant to Smolin's recent proposal that a form 
of natural selection operates on the level of universes, which favours 
the type of universe we live in. 
 } 
 \vfill 

 {\it Short Title:} Signature Change in Black Holes \\ 
 \vfill 

 \vfill 

 \end{center} 

 \newpage 

 \noindent {\large \bf 1.~ Introduction} 

  Space-time in general relativity is usually considered to possess a 
metric of Lorentzian signature. Positive definite metrics, with a 
Euclidean signature, have come into prominence lately through the 
Hartle and Hawking program concerning the wave function of the 
universe 
 \cite{H&H} --- \cite{Sakh}. 
 A general aim of that program is to try get a handle on the boundary 
conditions of the universe, and an intriguing suggestion made in 
\cite{NBC,H&H} is that the universe has no boundary, i.e. no origin 
where initial conditions have to be set, which is only possible if 
 space-time emerged from a Euclidean region preceeding a change of 
signature.  Another interesting development is the introduction of 
Euclidean wormholes. These wormholes can arise in one universe and 
connect it either to itself or to another universe. In order to 
attribute a transition probability, for example, between two 
Lorentzian regions, integration of the action along the tube 
connecting the two regions under study is required. In normal 
Lorentzian space the path integral approach leads to oscillating 
behavior, and hence to
 non-convergence of the integral. To obtain convergence, the 
substitution $t\rightarrow it$ is applied, thus introducing a 
Euclidean signature. This means in effect that we have two Lorentzian 
regions connected through a classically forbidden Euclidean region. 

Paralleling the Quantum cosmology program, papers 
 \cite{DMT1} --- \cite{SAH1} 
 pointed out that the possibility of a change in the signature of the 
metric is not restricted to a quantum description of General 
Relativity.  It was shown in \cite{rwsign,covar} that classical 
General Relativity does not prevent the existence of Euclidean regions 
and some examples of signature change in the 
 Friedmann-Lema\^{\i}tre-Robertson-Walker metric were produced.  
Further investigations were pursued in 
 \cite{KK1} --- \cite{HTTD}. 
 Even though the metric signature is invisible to the Einstein Field 
Equations, it should be noted \cite{rwsign,cont,CHTD2} that a change 
of signature is not, either $g_{00}$ goes through zero, in which case 
the metric is degenerate and $g^{00}$ singular there, or both $g_{00}$ 
and $g^{00}$ jump from positive to negative values, in which case the 
metric is discontinuous.  In either case, the Einstein Field equations 
cannot be defined in the usual way at the signature change 
\cite{CHTD2,TD}. 

  The 
 Friedmann-Lema\^{\i}tre-Robertson-Walker model has been considered 
repeatedly in both classical and quantum cosmological signature 
change, but few other models have been considered, especially in the 
classical signature change literature.  This paper constructs a 
classical model of signature change within the black hole topology, 
using Schwarzschild and 
 Lema\^{\i}tre-Tolman models 
 --- i.e. a Kruskal-Szekeres type of manifold. 
  It examines the transition from a black hole, through a signature 
change to a Euclidean region which reverses the collapse process, 
leading to a second signature change, and the birth of a white hole 
and a new universe.  We also determines whether the signature change 
surface can be hidden inside the horizon.  It continues the approach 
of papers \cite{rwsign,covar,cont} by exploring strictly classical 
signature changes in the Schwarzschild and 
 Lema\^{\i}tre-Tolman metrics. 

The investigation of transitions between Lorentzian geometries through 
a Euclidean region are also of interest when considered in conjunction 
with Smolin's idea \cite{smo}. Smolin's hypothesis is a proposed 
mechanism for determining the particular values of fundamental 
physical constants observed today, and thus justifying the anthropic 
principle. 
 In Smolin's paper, life supporting characteristics are linked to the 
existence of stars whose abundance is  linked to  the abundance of 
black holes. 
 It is envisaged that each `universe' either expands and
 re-collapses or expands indefinitely, possibly forming one or more 
black holes.  Instead of classical singularities occuring 
 --- a crunch singularity or black hole future singularities --- 
quantum cosmological tunneling gives birth to new universes, and hence 
a `natural history' of universes arises.  Furthermore, Smolin proposes 
that the process of tunnelling generates small random changes 
 --- `mutations' --- in the values of the physical constants.  Those 
combinations of values for which the universe generates many black 
holes will lead to large numbers of offspring having very similar 
values.  Thus, after the passage of many generations of universes, the 
population of universes will come to be dominated by those that 
generate lots of stars and black holes.  This parallels natural 
selection in that the `fittest' universes reproduce prolifically, but 
differs in that all blood lines (sets of constants) survive. 

  In what follows, we consider the junction conditions at a signature 
change in a Schwarzschild type metric, the choice of signature change 
surface, the form of the metric in the Euclidean region, and how 
geodesics should be propagated through.  Generalisations lead 
naturally to the 
 Lema\^{\i}tre-Tolman metric and its
 Kantowski-Sachs limit, which allow more interesting results. 

  The first step one has to take is to ensure that the various regions 
composing the space match geometrically.  We adopt the Darmois 
junction conditions \cite{dar}, and the application of them to 
signature change as presented in \cite{cont}. 

 \noindent {\large \bf 2.~ Junction Conditions and Conservation Laws} 

 \noindent {\it Conventions} 

We here work with 
 4-dimensional manifolds of Lorentzian and Euclidean signatures $(-
+++)$ and $(++++)$ respectively.  Greek indices range 
 0-3 and Latin indices 1-3.  Subscripts $E$ and $L$ denote quantities 
defined in or evaluated in Euclidean and Lorentzian regions 
respectively, and expressions without such subscripts are valid in 
either region.  These may also be written as superscripts, to avoid 
confusion with tensor indices.  Geometric units are used, $G = c = 1$, 
and the cosmological constant is set to zero, $\Lambda = 0$. 

 \noindent {\it Darmois Matching conditions} 

In standard Darmois matching, where no signature change occurs, space 
is composed of two regions, $V^+$ and $V^-$, with a common boundary 
surface $\Sigma$.  More precisely, an isomorphism $\psi : \Sigma^+ 
\rightarrow \Sigma^-$ allows us to identify the boundaries of $V^\pm$, 
$\Sigma^+ = \Sigma^- = \Sigma$ . The two regions have coordinate 
charts $x^\mu_+$ and $x^\mu_-$ and metrics $g^+_{\mu\nu}$ and $g^-
_{\mu\nu}$ respectively.  Setting the intrinsic coordinates of the 
junction surface to be $\xi^i_+ = \xi^i_- = \xi^i$, the locus of the 
surface is given parametrically in $V^\pm$ by $x^\mu_\pm = 
x^\mu_\pm(\xi^i)$, or by $\Xi^\pm(x^\mu_\pm) = 0$.  We write 
$Q\mid^\pm$ to denote evaluation of some quantity $Q$ in the limit as 
the surface is approached from either region, and $[Q]$ to denote the 
difference between the two limiting values 
 \begin{equation} 
    [Q] = Q\mid^+\;-\;Q\mid^-  \label{+--} 
 \end{equation} 
 The Darmois conditions \cite{dar} require the continuity of the first 
and second fundamental forms of the junction surface 
 --- i.e. the intrinsic metric and the extrinsic curvature.  The 
intrinsic metric is obtained by projecting the 
 4-metric onto $\Sigma$ using the basis vectors $e^\mu_i$ of $\Sigma$ 
 \begin{equation} 
    {}^3\!\!g_{ij} = g_{\mu\nu} e^\mu_i e^\nu_j \; , \;\;\;\; 
         e^\mu_i = \frac{\partial x^\mu}{\partial \xi^i} 
 \end{equation} 
 The extrinsic curvature describes the surface's shape in the 
enveloping space, and is the projection onto $\Sigma$, of the rate of 
change of the surface normal $n^\mu$ in the enveloping space, with 
respect to position on $\Sigma$. 
 \begin{equation} 
   K_{ij} = \left(\nabla_\mu n_\nu\right) e^\mu_i e^\nu_j = 
     -n_{\lambda} \left( \frac{\partial^2 x^\lambda}{\partial 
        \xi^i \partial \xi^j} + \Gamma^\lambda_{\mu\nu} 
        \frac{\partial x^\mu}{\partial \xi^i} 
        \frac{\partial x^\nu}{\partial \xi^j} \right) , 
   \;\;\;\;  n_{\nu} = \pm \frac{\partial_\nu \Xi} 
   {\sqrt{\epsilon_n \partial^\mu \Xi \partial _\mu \Xi}} 
         \label{normal} 
 \end{equation} 
 where $\epsilon_n = n^\nu n_\nu = +1$ if $\Sigma$ is 
 time-like and $-1$ if it is 
 space-like.  To conform with (\ref{+--}), the sign in (\ref{normal}) 
is set so that the $n^\nu_\pm$ point from $V^-$ to $V^+$ on both sides 
of $\Sigma$.  The Darmois conditions may now be given as: 
 \begin{equation}  [{}^3\!\!g_{ij}] = 0 \;\;\;\;\&\;\;\;\; 
    [K_{ij}] = 0 \end{equation} 
 In the constant signature case, according to \cite{bon}, these are 
equivalent to the Lichnerowicz matching conditions \cite{li}, whereas 
the O'Brien and Synge conditions \cite{os} are too restrictive. 

When we introduce a signature change at $\Sigma$, the equivalence 
between the Darmois and Lichnerowicz conditions breaks down.  Both the 
Lichnerowicz and O'Brien and Synge conditions insist that all the 
 4-d metric components be matched on either side of the junction 
surface, leading to a degenerate metric, a 
 non-affine time coordinate, and breakdown of the Einstein field 
equations.  We select the Darmois matching conditions as they are 
invariant to the coordinates chosen on either side.  They require no 
modification at a surface of signature change.  In fact they are blind 
to the change of signature, thus extending the signature blindness of 
the Einstein field equations.  A signature change surface is 
necessarily 
 space-like, so $\epsilon_n = n_\mu n^\mu = +1$ in the Euclidean 
region, and $-1$ in the Lorentzian region. 

 \noindent {\it Conservation Laws} 

In \cite{cont,cont2} the implications of signature change for 
conservation laws are worked out.  Conservation laws are based on the 
divergence theorem 
 --- i.e. the components version of Stokes theorem for a 
 3-form in a metric space.  The theorem requires a region $W$, bounded 
by a closed surface $S$ with outward pointing unit normal $m_\alpha$, 
smooth 
 non-zero volume elements $d^4W$ and $d^3S$ on $W$ and $S$ 
respectively, a smooth 
 non-zero metric, so that the inverse metric is well defined, and a 
smooth field $\Psi^\delta$: 
 \begin{equation} 
   \oint_S \Psi^\delta m_\delta d^3S = \int_W \nabla_\delta 
\Psi^\delta d^4W   \label{Div} 
 \end{equation} It should be noted that $n^\alpha$ is the normal to 
the junction surface $\Sigma$, and $m_\alpha$ is the normal to $S$, 
the closed boundary of $W$; 

 These conditions are not satisfied through a signature change.  Thus 
physical conservation laws need to be revised.  For the 
 electro-magnetic field we work with the 
 4-current $\Psi^\delta = J^\delta$, and for the gravitational field, 
a component of the Einstein tensor $\Psi^\delta = G^{\gamma \delta} 
v_\gamma$ where $v_\gamma$ is some suitable smooth vector field.  
Since $v_\gamma$ and $v^\gamma$ are not both smooth through a 
signature change, $\Psi^\delta = G^\delta_\gamma v^\gamma$ is also 
considered. 

Firstly, at a boundary where no signature change occurs, the Darmois 
junction conditions may be used to patch together two regions that 
adjoin the boundary on either side, and within which the divergence 
theorem does hold.  It is shown that these conditions, which give rise 
to Israel's identities \cite{Isr} for the Einstein tensor, 
 \begin{eqnarray} 
   [ G_{\mu\nu} n^\mu n^\nu ] &=& 0 \\ 
   {[ G_{\mu\nu} e^\mu_i n^\nu ]} &=& 0 
 \end{eqnarray} 
 where 
 \begin{eqnarray} 
   G_{\mu\nu} n^\mu n^\nu  &=&  
        \frac{1}{2} \{ K^2 - K_{ij} K^{ij} - \epsilon_n {}^3\!\!R \} 
                 \label{I1}  \\ 
   G_{\mu\nu} e^\mu_i n^\nu  &=&  {}^3\!\nabla_j K_i^j 
    - {}^3\!\nabla_i K  \label{I2} 
 \end{eqnarray} 
 ${}^3\!\!R$ and ${}^3\!\nabla_i$ being the intrinsic curvature 
invariant and covariant derivative of the 
 3-surface, and $K=g^{ij}K_{ij}=K^j_j$, are sufficient to ensure 
conservation of 
 energy-momentum through $\Sigma$.  Combined with suitable junction 
conditions on the 
 electro-magnetic field, they also ensure conservation of 
 4-current, with similar results applying to other fields. 

At a surface of signature change, $\epsilon_n$ now flips sign across 
$\Sigma$, and this leads to modified Israel identities 
 \begin{eqnarray} 
   [ G_{\mu\nu} n^\mu n^\nu ] & = & -{}^3\!\!R  \label{NI1}  \\ 
   {[ G_{\mu\nu} e^\mu_i n^\nu ]} & = & 0  \label{NI2} 
 \end{eqnarray} 
 It is necessary to distinguish two normals to $\Sigma$: $l_\delta = 
\partial \xi^0 / \partial x^\delta = \overline{e}^0_\delta$ and 
$n^\delta = \partial x^\delta / \partial \xi^0 = e^\delta_0$ where 
$n^\gamma n_\gamma = \epsilon_n = l_\gamma l^\gamma$ and $g_{\gamma 
\delta} n^\delta = \epsilon_n l_\gamma$.  A similar analysis of the 
divergence theorem through $\Sigma$ is made, paying careful attention 
to index position, the definition of the extrinsic curvature, and the 
directions of the various normals.  It is found that, in the process 
of patching together the two divergence theorems on either side of the 
signature change, the combined theorem aquires a surface term, so 
conservation laws must in general be modified.  The result is 
 \begin{equation} 
    \oint_{S} \Psi^\beta m_\beta \; d^3S 
     - \int_{S_o} E \; d^3S_o 
        = \int_{W} \nabla_\beta \Psi^\beta \; d^4W 
            \label{SCDiv} 
 \end{equation} 
 where 
 \begin{equation} 
  E = ( \Psi^\alpha_+ l_\alpha^+ - \Psi^\alpha_- l_\alpha^- ) 
          = [\Psi^\alpha l_\alpha]  \label{E} 
 \end{equation} 
 and $S_o$ is the region of $\Sigma$ enclosed by $S$.  For each of the 
four choices $\Psi^\delta = G^{\gamma \delta} v_\gamma$, $v_\gamma = 
l_\gamma$ or $\overline{e}^i_\gamma$ and $\Psi^\delta = 
G^\delta_\gamma v^\gamma$, $v^\gamma = n^\gamma$ or $e^\gamma_i$ the 
surface term $E = E(v_\gamma)$ or $E(v^\gamma)$ in the conservation 
law is 
 \begin{eqnarray} 
  E(l_\alpha) = [G^{\alpha \beta} l_\alpha l_\beta] & = & 
       - {}^3 \!\!R  \label{El}  \\ 
  E(\overline{e}^i_\alpha) = 
       [G^{\alpha \beta} l_\alpha \overline{e}^i_\beta] & = & 
       2({}^3\!\nabla_j K^{ij} - {}^3\!\!g^{ij} {}^3\!\nabla_j K) 
       \label{Eeb}  \\ 
  E(n^\alpha) = [G^\alpha_\beta l_\alpha n^\beta] & = & 
       K^2 - K_{ij} K^{ij}  \label{En}  \\ 
  E(e^\alpha_i) = [G^\alpha_\beta l_\alpha e_i^\beta] & = & 
       2({}^3 \! \nabla_j K_i^j - {}^3 \! \nabla_i K)  \label{Ee} 
 \end{eqnarray} 

The main results are expressed in a way that allows any set of 
junction conditions to be applied at the signature change, so this 
conclusion is independent of choice of junction conditions, as well as 
being coordinate invariant.  Alternative approaches to signature 
change, which emphasise maximum smoothness of the metric and the 
matter, are able to eliminate some, but not all of the surface 
effects.  Removal of all surface effects requires that the surface of 
signature change not only have zero extrinsic curvature, but also have 
zero 
 (3-d) Ricci scalar, which eliminates all realistic cosmological 
models. 

At a signature change then, the Darmois conditions still impose the 
same number of metric conditions as was sufficient for no signature 
change, and they still result in a modified set of conservation laws, 
albeit with surface effects.  These can be understood as a consequence 
of the change in the character of physical laws.  Whilst it is of 
interest to follow the maximum continuity, Lichnerowicz type approach, 
it is argued that physically interesting scenarios may be eliminated 
by it.  The following models are an example. 

 \noindent {\large \bf 3.~ The Schwarzschild Case} 

In the usual Schwarzschild line element \cite{sch} the signs of the 
metric components $g_{TT}$ and $g_{RR}$ interchange across $R=2M$, 
thus leading to reinterpertaion of the roles of $R$ and $T$.  
Consequently it is not clear which sign we should change to introduce 
the signature change, assuming that the general form of the metric is 
retained.  We shall investigate both possibilities, so we insert two 
new sign factors $\epsilon_T = \pm 1$ and $\epsilon_R = \pm 1$ in the 
metric: 
 \begin{equation} ds^2 = \epsilon_T \left( 1-\frac{2M}{R} \right) dT^2 
 + \epsilon_R \left(\frac{2M}{R}-1 \right)^{-1} dR^2 + R^2 d\Omega^2 
   \label{metric} 
 \end{equation} 
 where $d\Omega^2 = d\theta^2 + \sin^2{\theta}\,d\phi^2$.  We are 
interested only in transitions from the standard Schwarzschild metric 
to a Euclidean region, so we disregard the sign combination which 
gives us a second 
 (non-vacuum) Lorentzian manifold ($\epsilon_T = +1, \epsilon_R = 
+1$).  With both the other sign combinations ($\epsilon_T = -1, 
\epsilon_R = +1$ and $\epsilon_T = +1, \epsilon_R = -1$) it appears 
that there is a Euclidean region on one side of $R = 2M$, and a 
``double Lorentzian'' (or ``Kleinian") region (with two 
 time-like coordinates) on the other side.  It will become clear that 
the two Euclidean `regions' are in fact geodesically complete 
manifolds. 

Whilst Euclidean regions have no time, there will be a direction which 
is the extension of the time direction in the Lorentzian region.  One 
may determine whether the Euclidean metric is `static' or `dynamic' 
relative to this direction. 

The Einstein tensor components for this metric are \cite{GRTen} 
 \begin{equation} 
   G_{TT} = \frac{-\epsilon_T (1 + \epsilon_R) (1 - 2M/R)}{R^2}, 
\;\;\;\; 
   G_{RR} = \frac{(1 + \epsilon_R)}{R^2 (1 - 2M/R)}, \;\;\;\; 
   G_{\theta\theta} = 0, \;\;\;\; G_{\phi\phi} = 0 
 \end{equation} 
 A vacuum solution requires $\epsilon_R = -1$.  Transitions requiring 
a change of sign in $\epsilon_R$ introduce 
 non-vacuum solutions with strangely behaved matter, which we shall 
consider, since we don't really know what to expect in Euclidean 
signature physics.  The Ricci scalar and the Kretschmann scalar are 
 \begin{equation} 
   R^\mu_\mu = \frac{2(1 + \epsilon_R)}{R^2} , \;\;\;\; 
   k = R^{\mu\nu\lambda\sigma} R_{\mu\nu\lambda\sigma} = 
      \frac{8 [(1 + \epsilon_R) R^2 (1 - 2M/R) + 
             6 M^2]}{R^6}   \label{kret} 
 \end{equation} 
 where $R_{\mu\nu}$ and $R_{\mu\nu\lambda\sigma}$ are the Ricci and 
Riemann tensors.  Regardless of signature, $k$ is always singular at 
$R=0$,  and never at $R=2M$, so the Euclidean region $(\epsilon_T = 
+1, \epsilon_R = -1, R \geq 2M)$ has no singularities. 

We now investigate whether signature change is possible on two simple 
spacelike surfaces 
 --- in $R \geq 2M$ a constant $T$ surface, and in $R < 2M$ a constant 
$R$ surface.  More general transition surfaces will be considered 
using a different metric form.  The surface coordinates $\xi^i$ may 
then be chosen to be identically 3 of the enveloping coordinates 
$x^\mu$ in $V^-$ and $V^+$ 
 --- viz $(T, \theta, \phi)$ or $(R, \theta, \phi)$.  (c.f. \cite{Nag} 
in which the Euclidean solution for vacuum with a cosmological 
constant is found.) 

 \noindent {\it Constant $T$ surface} 

The intrinsic metric, unit normal, and extrinsic curvature of $\Sigma$ 
are given in either region by 
 \begin{eqnarray} 
   d\sigma^2 & = & \frac{-\epsilon_R}{(1 - 2M/R)} dR^2 + R^2 d\Omega^2 
   , \;\;\;\;\;\;\;\;\;\;\;\;\;\;\;\;  \epsilon_T \epsilon_n = 1 
         \label{tsur}  \\ 
   n^\mu  & = & \frac{\delta^\mu_T}{\sqrt{ 
         \epsilon_T \epsilon_n (1 - 2M/R)}} , \;\;\;\; 
        n^\mu n_\mu = \epsilon_n = \epsilon_T , \;\;\;\; 
   K_{ij} = 0 
 \end{eqnarray} 
 and the surface effects are all zero. 
 \begin{equation} 
   E(l_\alpha) = \frac{2(1 + \epsilon_R)}{R^2} = 0 , \;\;\;\; 
   E(\overline{e}_\alpha^i) = 0 , \;\;\;\; 
   E(n^\alpha) = 0 , \;\;\;\; 
   E(e^\alpha_i) = 0 
 \end{equation} 
 The two choices of future direction for $n^\mu$ are equivalent in a 
static metric.  The choice of a standard Schwarzschild solution in the 
Lorentzian region sets the sign of $\epsilon_R$ to $-1$, so 
$\epsilon_T$ flips across $\Sigma$, and this requires that $R_L \geq 
2M_L$ for a 
 space-like surface. Although (\ref{tsur}) is singular at 
$R_\pm=2M_\pm$, all constant $T$ surfaces on the Lorentzian side 
intersect that point, which is only a coordinate singularity, being 
the middle of the Schwarzschild wormhole at its moment of maximum 
expansion.  Applying $[{}^3\!\!g_{ij}] = 0$ and requiring the angular 
coordinates on either side to coincide, $\theta_E = \theta_L,\; \phi_E 
= \phi_L$, also fixes the areal radius and mass terms to be the same, 
$R_E = R_L$ and $M_E = M_L$.  Obviously $[K_{ij}] = 0$ imposes no 
further constraints.  Since all $T=$ constant surfaces are equivalent 
(for a static metric), this result is not surprising. 

This matching corresponds to vacuum both before and after the 
signature change, but the change surface extends to spatial infinity 
in both exterior regions $R_L > 2M_L$.  We are really seeking a change 
surface that is near the singularity $R=0$ and hidden inside the 
horizon.  No spacelike surface is further from $R = 0$ or less hidden 
than that of constant $T$. 

Since the middle of the throat at maximum expansion is a stationary 
point of the Killing vector $\chi^\mu = \delta^\mu_T$ on the 
Lorentzian side, and since all Lorentzian constant $T$ surfaces match 
to all Euclidean constant $T$ surfaces, this must be a stationary 
point on the Euclidean side also. This leads us to suspect that it is 
not possible to find two separate, 
 non-intersecting $T =$ constant surfaces in the Euclidean region.  In 
other words, we can't construct a Euclidean region between two 
separate Lorentzian regions. 

 \noindent {\it Constant $R$ surface} 

This is the simplest 
 non-vacuum case.  The fundamental forms and surface effects are: 
 \begin{eqnarray} 
   d\sigma^2 & = & \epsilon_T (1 - 2M/R) dT^2 + R^2 d\Omega^2 , 
     \;\;\;\;\;\;\;\;\;\;\;\;\;\;\;\;  \epsilon_R \epsilon_n = 1  \\ 
   n^\mu & = & -\sqrt{\epsilon_R \epsilon_n (2M/R - 1)} \; 
    \delta^\mu_R , \;\;\;\; n^\mu n_\mu = \epsilon_n = \epsilon_R  \\ 
   K_{TT} & = & \frac{- \epsilon_T M 
      \sqrt{\epsilon_R \epsilon_n (2M/R - 1)}}{R^2} ,   \nonumber  \\ 
   K_{\theta\theta} & = & - R \sqrt{\epsilon_R \epsilon_n (2M/R - 1)} 
         , \;\;\;\; 
   K_{\phi\phi} = \sin^2\theta \; K_{\theta\theta}  \\ 
   E(l_\alpha) & = & \frac{-2}{R^2} , \;\;\;\; 
   E(\overline{e}_\alpha^i) = 0 , \;\;\;\; 
   E(n^\alpha) = \frac{-2 \epsilon_R \epsilon_n}{R^2} 
       = \frac{-2}{R^2} , \;\;\;\; 
   E(e^\alpha_i) = 0 
 \end{eqnarray} 
 where we have chosen $n^\mu$ to point in the direction of collapse, 
i.e. towards $R$ decreasing.  A similar analysis gives us $\epsilon_T 
= -1$, $R_L < 2M_L$, and $\epsilon_R$ flips; choosing $\theta_E = 
\theta_L, \; \phi_E = \phi_L$ and $T_E = T_L$ with $[{}^3\!\!g_{ij}] = 
0 \Rightarrow R_E = R_L$, and $M_E = M_L$.  No further restrictions 
are necessary to ensure $[K_{ij}] = 0$. 

This demonstrates that matching can be achieved on a surface that is 
entirely inside the horizon, but since the Euclidean region is not 
empty in this case, it still leaves open the interpretation of the 
energy stress tensor on the Euclidean side.  Further, since $R$ is the 
timelike coordinate on the Lorentzian side, $R$ is the nominal `time' 
direction on the Euclidean side too, being orthogonal to the 
transition surface, so this Euclidean manifold, with $R < 2M$, is 
`dynamic'. 

 \noindent {\it Geodesic Coverage} 

What do the particle paths look like in  the combined space?  We now 
investigate the behavior of radial timelike geodesics.  Adding the 
angular components of the motion should present no problem, as 
$\theta_\pm$ and $\phi_\pm$ are identified at $\Sigma$.  One aim is to 
verify that the space is geodesically complete, and in the context of 
this paper geodesics that end on a curvature singularity are 
considered to be as complete as is possible.  The second aim is to see 
how geodesics should be continued at the transition, and whether the 
set of all geodesics arriving at the Lorentzian side of the transition 
generate all possible geodesics emerging on the Euclidean side.  Three 
schemes for continuing geodesics are considered.  Two of them attempt 
to match particle 
 4-velocities (unit normal tangent vectors), and one attempts to match 
 4-momenta. 

The geodesic equation 
 \begin{equation}   u^\mu \nabla_\mu u^\nu = 0   \end{equation} with 
the condition for a
 `time-like' unit normal 
 \begin{equation}   u^\mu u_\mu = \epsilon_n   \label{t-l-norm}
\end{equation} 
 where $\epsilon_n = \epsilon_T$ at a constant $T$ transition or 
$\epsilon_R$ at a constant $R$ transition, leads to the acceleration 
 \begin{equation}  \ddot{R} = \frac{-\epsilon_R \epsilon_n M}{R^2} \;, 
    \end{equation} 
 and gives the following unit tangent vectors, where the signs have 
all been chosen so that positive $h$ and $q$ values always give 
consistently future directed infalling tangent vectors where $T > 0$. 
 \begin{equation}  \mbox{Lorentzian:~~~~~~~~~~~~~} u^\mu = \left( 
    \frac{h_L}{(1-2M/R)} \;, - q_L \sqrt{h_L^2 - (1 - 2M/R)}\;, 0, 0 
\right) , 
     \;\;\;\; q_L = \pm 1  \end{equation} 
 There are three types of geodesics: ~(a) $1-2M/R \leq h^2_L < 1 $: 
geodesics recollapsing from past to future singularities, $R = 0$, 
with a maximum at $R=2M/(1-h^2_L) \geq 2M$; ~(b) $h^2_L>1$:  
monotonically ingoing or outgoing geodesics with finite velocity 
$\sqrt{h_L^2 - 1}$ at $R=\infty$, reaching $R = 0$ either in the past 
or future; ~(c) $h^2_L = 1$: marginal monotonic geodesics with zero 
velocity at $R=\infty$. 
 \begin{equation}  \mbox{Euclidean, $R \geq 2M$:~~~~~~} u^\mu = \left( 
    \frac{h_E}{(1-2M/R)} \;, - q_E \sqrt{(1 - 2M/R) - h_E^2}\;, 0, 0 
\right) , 
     \;\;\;\; q_E = \pm 1  \end{equation} 
 In this case there is only one type of geodesic, descending from $R = 
\infty$, through a minimum at $R = 2M/(1-h_E^2) \geq 2M$, and 
 re-expanding back out.  The allowed range of $h_E$ is then $0 \leq 
h_E^2 \leq (1-2M/R)$, and all geodesic paths are restricted to the 
region $R \geq 2M$. The only geodesic reaching $R=2M$ is the one with 
$h_E=0$ which in effect is a stationary point. This is in accord with 
the range of $R$ at $\Sigma$, and confirms that the region $R \geq 2M$ 
is a geodesically complete manifold. 
 \begin{equation}  \mbox{Euclidean, $R < 2M$:~~~~~~} u^\mu = \left( 
    \frac{-h_E}{2M/R-1} \;, - q_E \sqrt{(2M/R - 1) - h^2_E}\;, 0, 0 
\right) , 
     \;\;\;\; q_E = \pm 1  \end{equation} 
 These geodesics all expand from $R=0$ and 
 re-collapse back, and have maxima at $R = 2M/(1 + h_E^2) \leq 2M$, 
where $h_E^2 \leq (2M/R - 1)$.  Thus the region $R \leq 2M$ is also 
geodesically complete, but does not have the desired bouncing 
property.  ($R = 2M$ would not actually be encountered in this region 
of our model, since $R_\Sigma < 2M$.) 

Since geodesic tangent vectors have a unit magnitude that flips sign 
across a signature change, it is impossible to match all components of 
$u^\mu_\pm$ across $\Sigma$.  We consider here three possible schemes, 
and summarise the resulting conditions in Table \ref{GeodCont} below: 
\\ 
 \hspace*{7mm}(i) Match $u^R$ values, with $u^\mu u_\mu = \epsilon_n$ 
giving a jump in the $u^T$ values; \\ 
 \hspace*{7mm}(ii) Match $u^T$ values, with $u^\mu u_\mu = \epsilon_n$ 
giving a jump in the $u^R$ values; \\ 
 \hspace*{7mm}(iii) \parbox[t]{14cm}{Match all components of the 
 4-momentum $P^\mu = m u^\mu$, allowing $P_\mu$ and $|P^\mu P_\mu| = 
m^2$ to jump.  (This is equivalent to matching 
 non-normalised tangent vectors, for which the metric degeneracy at 
$\Sigma$ is irrelevant, and standard existence and uniqueness theorems 
guarantee that geodesic continuation of all tangent vectors is 
possible.)} 

 \begin{center} 
               ======================= \\ 
                  TABLE 1 GOES HERE \\ 
               ======================= 
 \end{center} 

 \begin{center} 
               ======================== \\ 
                  TABLE 2  GOES HERE \\ 
               ======================== 
 \end{center} 

 When these are compared with the allowed ranges of $h$ in each 
region, summarised in Table \ref{hRanges}, we see that conditions (i) 
and (ii) do not allow the continuation of all possible geodesics that 
might arrive at either type of signature change surface, whereas (iii) 
continues all geodesics at both types.  For example, identifying 
$u^R_E = u^R_L$ at a constant $T$ surface gives us 
 \begin{equation} 
   h_E^2 = 2(1 - 2M/R) - h_L^2 , \;\;\;\; R \geq 2M , \;\;\;\; 
   h_L^2 \geq (1 - 2M/R) , \;\;\;\; h_E^2 \leq (1 - 2M/R) 
 \end{equation} 
 so we can easily find a large enough value for $h_L^2$ to make $h_E$ 
imaginary.  For case (iii) we find 
 \begin{eqnarray} 
   \mbox{Constant $T$ transition:~~~~~~} & & 
   \left(\frac{m_E}{m_L}\right)^2 = \left(\frac{h_L}{h_E}\right)^2 
         = \frac{2 h_L^2}{(1 - 2M/R)} - 1    \label{massratioT} \\ 
   \mbox{Constant $R$ transition:~~~~~~} & & 
   \left(\frac{m_E}{m_L}\right)^2 = \left(\frac{h_L}{h_E}\right)^2 
         = \frac{2 h_L^2}{(2M/R - 1)} + 1    \label{massratioR} 
 \end{eqnarray} 
 Since $h_L^2 \geq (1 - 2M/R)$ the particle's Euclidean mass is always 
greater than it's Lorentzian mass $m_E \geq m_L$, as well as $h_L^2 
\geq h_E^2$. 

 Thus it turns out that condition (\ref{t-l-norm}), forcing the 
tangent vectors to be unit vectors, is too strong an assumption, and 
doesn't permit all particle paths to be matched through $\Sigma$.  
Rather, the matching of geodesic 
 4-momenta is the only way of extending all particle paths through a 
signature change.  The conclusion that the 
 `rest-mass' parameter of a particle has to jump, is consistent with 
the fact \cite{cont} that in general tensors cannot be smooth through 
$\Sigma$ in both covariant and contravariant forms, and that the 
density can jump across a signature change.  In Fig. 1 we summarise 
the properties of geodesics arriving at $\Sigma$ at some particular 
value of $R_\Sigma$.  The horizontal axis is the parameter $h_L^2$ and 
the plot covers a representative range of permissible $h_L^2$ values 
for the chosen $R_\Sigma$. 

 \begin{center} 
               ====================== \\ 
                  FIG 1  GOES HERE \\ 
               ====================== 
 \end{center} 

 \noindent {\it Summary} 

Within the Schwarzschild metric form, signature change is possible on 
constant $R$ surfaces inside the horizon, but the resulting Euclidean 
region has strange matter, and continues to collapse to a singularity.  
A second signature change back to a Lorentzian region is of course 
possible, but only at a smaller $R$, closer to the future singularity.  
Signature change is also possible on constant $T$ surfaces, leading 
into a Euclidean region which is vacuum and has geodesics which do 
bounce.  But constant $T$ surfaces are entirely outside the horizon, 
and so not very interesting, since a second transition back to a 
Lorenzian spacetime results in the same future as the one that was 
avoided. 

Attempts to continue geodesics through a signature change indicated 
that one must match the 
 4-momenta, which means that particle rest masses have to jump.  If a 
second signature change back to a Lorentzian metric occurred, the 
particle mass would not return to its original mass unless the model 
were highly symmetric. 

 \noindent {\large \bf 4.~ The Lema\^{\i}tre-Tolman Case} 

We now shift our attention to the 
 Lema\^{\i}tre-Tolman metric \cite{lem,tol}, primarily because it 
allows us to deal simply with more general surfaces in spherical 
vacuum, and secondly because it makes possible a generalisation of the 
black hole topology to 
 non-empty models, thus describing more realistically the collapse of 
matter into black holes, as well as the more standard cosmological 
collapse of matter, where no wormhole topology is involved \cite{ch2}.  
This gets us closer to Smolin's scenario. 

In the vacuum case, the 
 Lema\^{\i}tre-Tolman metric with appropriate choice of parameters can 
describe the full 
 Schwarzschild-Kruskal-Szekeres manifold, avoiding a coordinate 
singularity at $R = 2M$ and the accompanying change of character of 
the Schwarzschild $R$ and $T$ coordinates, and making it clear which 
metric element should change sign at a change of signature.  Its two 
arbitrary functions make it much more flexible than the 
 Kruskal-Szekeres metric. 

The diagonal, synchronous, spherically symmetric metric, with an added 
factor of $\epsilon = \pm1$, 
 \begin{equation} ds^2 = \epsilon dt^2 + B^2(t,r)dr^2 + 
R^2(t,r)d\Omega^2 \label{tmetric} \end{equation} 
 leads to the following Einstein tensor: 
 \begin{eqnarray} 
 G_{tt} &=& \frac{\epsilon (2BRR'' + BR'^2 - 2B'RR' - B^3) + 
   (2B^2\dot{B}R\dot{R} + B^3\dot{R}^2)}{B^3R^2}\\ 
 G_{tr} &=& \frac{2(\dot{B}R' - B\dot{R}')}{BR}\\ 
 G_{rr} &=& \frac{(R'^2 - B^2) + \epsilon(2B^2R\ddot{R} + 
        B^2\dot{R}^2)}{R^2}\\ 
 \frac{G_{\phi\phi}}{\sin^2 \theta} = G_{\theta\theta} &=& 
          \frac{R[(BR'' - B'R') + \epsilon(B^2\ddot{B}R + 
          B^2\dot{B}\dot{R} + B^3\ddot{R})]}{B^3}\\ 
 \end{eqnarray} 
 where $\; '\equiv \partial/\partial r \;\;\;\; \& \;\;\;\; \dot{ 
}\equiv \partial/\partial t \;$ and the cosmological constant is taken 
to be zero. 

Solving the Einstein field equations for 
 co-moving matter, $u^\mu = \delta^\mu_t$, and zero pressure, $p = 0$, 
gives the 
 Lema\^{\i}tre-Tolman model, and we get 
 \begin{eqnarray} 
   B^2 & = & \frac{(R')^2}{1 + f} , \;\;\;\;  f(r) \geq -1  
\label{tolgrr} \\ 
   -\epsilon \dot{R}^2 & = & \frac{2 M}{R} + f(r)  \label{tolvel} \\ 
   \ddot{R} & = & \epsilon \frac{M}{R^2}  \label{tolacc} \\ 
   8 \pi \rho = G_{tt} & = & -\epsilon \frac{2 M'}{R^2 R'}  
\label{tolden} \\ 
  R^\mu_\mu & = & 2 \left( 
    \left( \frac{2 B' R'}{B^3 R} - \frac{(R')^2}{B^2 R^2} 
                 - \frac{2 R''}{B^2 R}  \right) 
    + \frac{(1 - \epsilon \dot{R}^2)}{R^2} 
    - \epsilon \left( \frac{2 \ddot{R}}{R} 
           + \frac{2 \dot{B} \dot{R}}{B R} + \frac{\ddot{B}}{B} 
\right) 
        \right)  \nonumber  \\ 
         & = & \frac{2 M'}{R^2 R'} \\ 
   k & = & R^{\mu\nu\lambda\sigma} R_{\mu\nu\lambda\sigma} 
                     \nonumber   \\ 
     & = & 4 \left( \frac{2 \ddot{R}^2}{R^2} + \frac{\ddot{B}^2}{B^2} 
    + \left( \frac{1 - \epsilon \dot{R}^2}{R^2} 
          - \frac{(R')^2}{B^2 R^2} \right)^2 \right.  \nonumber  \\ 
    && \left. + 2 \left( \frac{R''}{B^2 R} - \frac{B' R'}{B^3 R} 
                    + \frac{\epsilon \dot{B} \dot{R}}{B R} \right)^2 
    + 4 \epsilon \left( \frac{\dot{R}'}{B R} 
                      - \frac{\dot{B} R'}{B^2 R} \right)^2 \right) 
           \nonumber  \\ 
   & = & 4 \left( \frac{3M'^2}{R^4R'^2} - \frac{8M'M}{R^5R'} + 
\frac{12M^2}{R^6} \right)  \label{tolkret} 
 \end{eqnarray} 
 where $f = f(r)$ and $M = M(r)$ are arbitrary functions of coordinate 
radius $r$, $\rho$ is the density, and $k$ is the Kretschmann scalar.  
Singularitites in $k$ and $\rho$ occur at $R= 0$ and $R' = 0$ 
regardless of $\epsilon$.  Shell crossings occur where $R' = 0$, since 
shells of matter at a different constant $r$, arrive at the same areal 
radius $R(t,r)$ and intersect each other.  In vacuum, $M' = 0$, so 
$\rho$ is zero and $k$ is finite, and there is no physical problem, 
but there is a bad coordinate coverage of the space. In 
 non-vacuum cases care needs to be taken to select the arbitrary 
functions which do not give rise to these physically troublesome shell 
crossings \cite{shcr}. 

In the standard Lorentzian case ($\epsilon = -1$), $f(r)$ is a kind of 
local energy constant which determines the type of time evolution 
 --- elliptic, parabolic or hyperbolic --- as well as the local 
geometry, and $M(r)$ represents the total effective gravitational mass 
within comoving radius $r$. 

In the Euclidean case ($\epsilon = +1$) the acceleration 
(\ref{tolacc}) is everywhere positive, provided we select a positive 
`mass' term, so a bouncing Euclidean universe can be achieved.  This 
requires $f_E$ to be negative in order to keep $\dot{R}$ real, i.e. 
 $-1 \leq f_E < 0$. 

We now obtain solutions to the evolution equation (\ref{tolvel}), in 
terms of a parameter $\eta$, and $a = a(r)$, a third arbitrary 
function of $r$, which is the time of the big bang $R = 0$, or if we 
use the time reverse of the following equations, the time of the big 
crunch.  Solutions with $f \geq 0$ or $M \leq 0$ are discarded.  
Although we can't be sure negative `mass' solutions are physically 
disallowed in a Euclidean manifold, they all reach the $R = 0$ 
singularity, and none of them bounce 
 (re-expand), so they do not serve our purpose. 

Lorentzian region $\epsilon  = -1$:~~~~elliptic solution, $-1 \leq f_L 
< 0$ 
 \begin{equation} 
   R(t,r) = \frac{M_L}{(- f_L)} (1 - \cos\eta_L) , \;\;\;\; 
   t = \frac{M_L}{(- f_L)^{3/2}} (\eta_L - \sin\eta_L) + a_L(r) 
  \label{lneq}   \end{equation} Euclidean region $\epsilon  = 
+1$:~~~~$M_E > 0$, $-1 \leq f_E < 0$ 
 \begin{equation} 
   R(t,r) = \frac{M_E}{(- f_E)} (\cosh\eta_E + 1) , \;\;\;\; 
   t = \frac{M_E}{(- f_E)^{3/2}} (\sinh\eta_E + \eta_E) + a_E(r) 
 \label{enf}  \end{equation} 

Any 
 Lema\^{\i}tre-Tolman model with $M'=0$ is a vacuum model, and thus 
for 
 $\epsilon = -1$ represents at least a section of the 
 Kruskal-Szekeres-Schwarzschild space time in geodesic coordinates.  
However not every selection of the arbitrary functions gives complete 
coverage of the manifold.  Novikov coordinates \cite{nov} do cover the 
entire manifold, and are obtained with the following choices 
 \begin{equation} 
    M_L = \mbox{const} , \;\;\;\; 
    f = \frac{-1}{1 + (r/2M_L)^2} , \;\;\;\; 
    a_L(r) = \frac{-\pi M_L}{(-f)^{3/2}}  \label{symmetry} 
 \end{equation} 
 for which the surface $t = 0$ is a simultaneous time of maximum 
expansion, and $f(0) = -1$ at the Schwarzschild throat, increasing 
monotonically to 0 as $r \rightarrow \pm\infty$.  This topology 
 --- two sheets joined by a throat 
 --- may easily be extended to 
 non-vacuum everywhere \cite{ch2} by setting $M_L = M_L(r)$ with a 
minimum value at the throat.  It is the form of $f(r)$ which 
determines the topology.  If the asymptotic regions are closed FLRW 
cosmologies ($f = -kr^2$, $k = +1$), then we still expect $f$ to rise 
very close to zero before decreasing again.  In such dense black 
holes, the past and future event horizons are split, and $R = 2M_L$ is 
an apparent horizon \cite{ch2}. 

 \noindent {\it Matching conditions} 

We perform the matching on the simplest possible surface, that of 
constant time, $t =$ constant. In vacuum this is merely a coordinate 
restriction and not a physical one, because the origin of the time 
coordinate, $a\left(r\right)$, is an arbitrary function of position. 
It amounts to finding the family of geodesics orthogonal to the 
transition surface, and using these as lines of constant $r$. The 
intrinsic metric of such a surface is correspondingly simple: 
 \begin{equation} d\sigma^2 = B^2 \, dr^2 + R^2 \, d\Omega^2 
    = \frac{(R')^2}{1 + f} \, dr^2 + R^2 \, d\Omega^2  \end{equation} 
 When matching, a reasonable choice is to equate the angular parts, 
and to 
 re-scale the coordinate radii, so that 
 \begin{equation}  \theta_E = \theta_L , \;\;\;\; \phi_E = \phi_L , 
\;\;\;\; 
        r_E = r_L \end{equation} 
 and $[{}^3\!\!g_{ij}] = 0$ fixes 
 \begin{equation}  R_E = R_L = R_\Sigma , \;\;\;\; 
    B_E = B_L = B_\Sigma   \end{equation} 
 Since $R$ is continuous across the junction and is a function of $r$ 
only on $\Sigma$, i.e. $R_\Sigma = R_\Sigma(r)$, we have also that 
 \begin{equation}  R'_E = R'_L = R'_\Sigma, \;\;\;\; 
   \Rightarrow f_E = f_L = f  \end{equation} 
 Because of (\ref{tolgrr}) the normal and the 
 non-zero elements of the extrinsic curvature are: 
 \begin{eqnarray} 
   n^\mu = \sqrt{\epsilon \epsilon_n} \, \delta^\mu_t , && \;\; 
       n^\mu n_\mu = \epsilon_n = \epsilon ,  \\ 
   K_{rr} = \sqrt{\epsilon \epsilon_n} \, B \dot{B} 
               = \frac{R' \dot{R}'}{1 + f} , && \;\;\;\; 
   K_{\theta\theta} = \sqrt{\epsilon \epsilon_n} \, R \dot{R} 
               = R \dot{R} = \frac{K_{\phi\phi}}{\sin^2\theta} 
 \end{eqnarray} 
 and $[K_{ij}] = 0$ leads to 
 \begin{equation} \dot{R}_L = \dot{R}_E , \;\;\;\; 
   \dot{B}_L = \dot{B}_E \;\; \Rightarrow \;\; 
    \dot{R}'_L = \dot{R}'_E  \label{Rdotel}  \end{equation} 
 The surface effects are 
 \begin{eqnarray} 
   E(l_\alpha) & = & 2 \left( 
      \frac{(R')^2}{B^2 R^2} - \frac{2 B' R'}{B^3 R} 
    + \frac{2 R''}{B^2 R} - \frac{1}{R^2} 
                  \right) 
        = 2 \left( \frac{f}{R^2} + \frac{f'}{R R'} \right)  \\ 
    B^2 E(\overline{e}_\alpha^r) = E(e^\alpha_r) & = & 
       4 \sqrt{\epsilon \epsilon_n} \, \left( 
       \frac{8 \dot{R} R'}{R^2} - \frac{9 \dot{R}'}{R} \right)  \\ 
    E(n^\alpha) & = & \frac{2 \epsilon \epsilon_n \dot{R}^2}{R^2} 
 \end{eqnarray} 
 The principal feature we are looking for is a bouncing universe, 
meaning a Lorentzian region matched to a bouncing Euclidean region 
that in turn may be  matched to another Lorentzian region. This 
involves establishing the existence of at least two solution surfaces 
in the Euclidean region of the model under investigation.  In general, 
given two 
 space-like hypersurfaces, there will not be any geodesics that are 
orthogonal to both, so requiring both to be $t=$ constant surfaces in 
the same coordinate system could well be restrictive. 

 \noindent {\it General transitions} 

Five arbitrary functions, $f(r)$, $M_E(r)$, $M_L(r)$, $a_L(r)$ and 
$a_E(r)$,  are as yet unspecified.  We now derive the necessary 
relations between them at a surface of signature change.  We do not 
assume vacuum at this stage.  Only models with $f < 0$ and $M_E > 0$ 
give rise to a Euclidean region with a bounce. 

Condition (\ref{Rdotel}) for $\dot{R}^2$ combines with the evolution 
equation (\ref{tolvel}) to give 
 \begin{equation}  M_L - M_E = \dot{R}^2_{\Sigma} R_{\Sigma} 
     \mbox{~~~~~~or~~~~~~} M_L + M_E = -fR_{\Sigma} \label{con} 
        \end{equation} 
 The sign of $\dot{R}$ must still be matched. 

 Inserting the parametric expressions for $R_\Sigma$ (\ref{lneq}) and 
(\ref{enf}) into (\ref{con}) gives 
 \begin{eqnarray} 
  \cos\eta_{L\Sigma} = - \frac{M_E}{M_L} , & \;\;\;\; & 
  \cosh\eta_{E\Sigma} =  \frac{M_L}{M_E}  \label{etal} \\ 
  \Rightarrow \;\;\;\; M_E \leq M_L \;\; , \;\;\;\;\;\;\;\; 
    -1 \leq & \cos\eta_L & \leq 0 \;\;, \;\;\;\;\;\;\;\; 
    1 \leq \cosh\eta_E \leq \infty 
 \end{eqnarray} 
 which, combined with the continuity of $\dot{R}$ for collapsing 
models, yields 
 \begin{equation} 
   t_\Sigma < 0: \;\;\;\; \eta_{L\Sigma} = 2\pi-\cos^{-1} (-M_E/M_L) , 
       \;\;\;\; \eta_{E\Sigma}= -\cosh^{-1} (M_L/M_E) \label{tneg} 
 \end{equation} 
 Since the Euclidean region doesn't have arbitrarily large $\dot{R}$, 
the transition cannot happen arbitrarily close to $R = 0$.  On the 
transition surface $t_\Sigma$ is constant, so $\eta_{L\Sigma}$ and 
$\eta_{E\Sigma}$ are functions of $r$ only. Thus 
 \begin{eqnarray} 
   t_{E\Sigma} = t_\Sigma & = & a_E(r) + \frac{M_E}{(- f)^{3/2}} 
         (\sinh\eta_{E\Sigma} + \eta_{E\Sigma})  \label{symEuc} \\ 
   t_{L\Sigma} = t_\Sigma & = & a_L(r) + \frac{M_L}{(- f)^{3/2}} 
         (\eta_{L\Sigma} - \sin\eta_{L\Sigma})  \label{algen} 
 \end{eqnarray} 
 If possible, we choose $a_E(r) = 0 $, so that (\ref{symEuc}) gives a 
 time-symmetric coordinate coverage in the Euclidean region. This 
permits a second copy of any transition surface found away from $t=0$, 
and thus ensures a bounce.  To obtain a specific solution, we fix 
$t_\Sigma$ and any two of $M_L, M_E, a_L, a_E$, to obtain the others.  

Further, (\ref{con}) plus the requirement that the transition surface 
be inside the external horizon $R = 2 M_{L\,max}$, where $M_{L\,max}$ 
is the total exterior mass of the collapsing cloud, gives us 
 \begin{equation} 
   f(r) < - \frac{M_L(r) + M_E(r)}{2M_{L\,max}} 
    <  - \frac{M_L(r)}{2M_{L\,max}}  \label{flmR2M} 
 \end{equation} 
 since $0 \leq M_E \leq M_L$.  At the centre we have 
 $f(0) < -(M_{L\,min}/2M_{L\,max})$ and outside the cloud or at large 
$r$, $f(r>r_{max}) < -(1/2)$, where $r_{max}$ is the smallest radius 
for which $M_L(r) = M_{L\,max}$.  This is a very stong restriction on 
$f$. 

 \noindent {\it Vacuum to Vacuum} 

We set the mass term constant (hence $\rho=0$) and equal on either 
side $M_L = M_E$.  By (\ref{con}) this gives $\dot{R}_\Sigma = 0$, and 
by (\ref{lneq}) and (\ref{enf}) the areal radius can only be matched 
at 
 \begin{equation} \eta_{E\Sigma} = 0 , \;\;\;\; \eta_{L\Sigma} = \pi , 
   \;\;\;\; R_\Sigma = \frac{2M}{(- f)} , \;\;\;\; 
     -1 \leq f \leq 0  \end{equation} 
 the loci of minimum and maximum expansion of the Euclidean and 
Lorentzian coordinates respectively.  Clearly, this case is equivalent 
to a constant $T$ transition in the Schwarzschild case, as $\Sigma$ 
touches $R=2M$ but otherwise lies entirely outside the horizon.  If we 
set $a_E = 0$, to obtain a symmetric coverage of the Euclidean region, 
we have that the transition time is $t_\Sigma = 0$ 
 --- i.e. there is no `time' between the two transitions.  This 
confirms our earlier suspicion that minimum expansion at the middle of 
the throat is also a unique event in the Euclidean Schwarzschild 
topology. 

Retaining $M_E$ and $M_L$ constant, but not necessarily equal, 
(\ref{etal}) shows that both $\eta_E$ and $\eta_L$ are constant on the 
transition surface.  The parametric expressions for $t$ 
(\ref{symEuc})-(\ref{algen}) then establish the relation between 
$f(r)$ and $a(r)$.  Again symmetric coverage of the Euclidean region 
 --- $a_E = 0$ in (\ref{symEuc}) --- would require $f =$ constant and 
hence $R_\Sigma =$ constant.  The only way to get surfaces which have 
constant $t$, $R$, and $f$, is to set $f = -1$ --- dealt with next. 

 \noindent {\it Constant $R$} 

The constant $t$ surfaces can also be made constant $R$ surfaces, thus 
yielding the closed 
 Kantowski-Sachs model \cite{KanSac} in 
 Lema\^{\i}tre-Tolman coordinates \cite{TKS}.  This is done by setting 
 \begin{equation} 
   M = M_1 \int \sqrt{1 + f} \; dr + M_0 , \;\;\;\; 
   a = a_1 \int \sqrt{1 + f} \; dr + a_0 , \;\;\;\; 
   M_0, M_1, a_0, a_1 \mbox{~constants} 
 \end{equation} 
 and then taking the limit $f \rightarrow -1$, leading to 
\begin{eqnarray} 
   8 \pi \rho & = & -\epsilon \frac{2 M_1}{R^2 B} \\ 
   \epsilon = -1: \;\;\;\; 
      R_L & = & M_{0L}(1 - \cos\eta_L) , \;\;\;\; 
      B_L = 2M_{1L} - (M_{1L}\eta_L + a_{1L}) 
                   \frac{\sin\eta_L}{(1 - \cos\eta_L)} \\ 
   \epsilon = +1: \;\;\;\; 
      R_E & = & M_{0E}(\cosh\eta_E + 1) , \;\;\;\; 
      B_E = 2M_{1E} - (M_{1E}\eta_E + a_{1E}) 
                   \frac{\sinh\eta_E}{(\cosh\eta_E + 1)} 
 \end{eqnarray} 
 In Lorentzian vacuum, $M_1 = 0, \epsilon = -1$, these coordinates 
only cover $R \leq 2M_L$, and may be similarly incomplete in dense 
models.  Shell crossings may be avoided for $-2\pi < (a_{1L}/M_{1L}) < 
0$ in Lorentzian regions, but not in Euclidean regions.  However, for 
$a_{1E} = 0$, shell crossings are removed if transitions happen at 
$\mid \eta_{E\Sigma} \mid < 2.3994$ which is the positive root of 
$2(\cosh\eta_{E} + 1) - \eta_E \sinh\eta_E = 0$. 

By (\ref{con}), $R_\Sigma = M_{0E} + M_{0L}$ and for $M_{0E} > 0$ and 
$\Sigma$ hidden ($R_\Sigma \leq 2M_{0L}$) we need $M_{0L} < R_\Sigma 
\leq 2M_{0L} \Rightarrow 0 < M_{0E} \leq M_{0L}$.  Equations 
(\ref{con})-(\ref{etal}) and 
 (\ref{symEuc})-(\ref{algen}) become 
 \begin{eqnarray} 
  & R_\Sigma = M_{0E} + M_{0L}, \;\;\;\; 
  \cos\eta_{L\Sigma} = - \frac{M_{0E}}{M_{0L}} , \;\;\;\; 
  \cosh\eta_{E\Sigma} =  \frac{M_{0L}}{M_{0E}} &  \\ 
   & a_{0E} + M_{0E} (\sinh\eta_{E\Sigma} + \eta_{E\Sigma}) 
   = t_\Sigma = a_{0L} + M_{0L} (\eta_{L\Sigma} - \sin\eta_{L\Sigma}) 
& 
 \end{eqnarray} 
 and because the matching of $R_\Sigma$ and $\dot{R}_\Sigma$ no longer 
ensures $B_\Sigma$ and $\dot{B}_\Sigma$ is matched, we get two extra 
conditions 
 \begin{eqnarray} 
   & (M_{0E}M_{1L} - M_{0L}M_{1E}) R_\Sigma + &  \nonumber  \\ 
   & ((M_{1L} - M_{1E}) t_\Sigma 
   - a_{0L}M_{1L} + a_{0E}M_{1E} + a_{1L}M_{0L} - a_{1E}M_{0E}) 
   \sqrt{M^2_{0L} - M^2_{0E}} = 0 & \\ 
   & (M_{1L} + M_{1E}) t_\Sigma 
   - a_{0L}M_{1L} - a_{0E}M_{1E} + a_{1L}M_{0L} + a_{1E}M_{0E}) = 0 & 
 \end{eqnarray} 
 Simplifications are obtained by requiring a symmetric Euclidean 
region $a_{0E} = 0$, $a_{1E} = 0$.  Vacuum to vacuum is not possible 
since $M_{1L} = 0$, $M_{1E} = 0$ implies $a_{1E} = 0 = a_{1L}$, and 
thus $B_L = 0 = B_E$ at all times.  Similarly, vacuum to 
 non-vacuum is not possible.  The dense models are highly symmetric, 
as they have uniform density on constant $R$ surfaces.  A sample set 
of values are:  $M_{0E}/M_{0L} = 0.1, \; \eta_{E\Sigma} = -2.9932, \; 
\eta_{L\Sigma} = 4.6122, \; a_{0E} = 0, \; t_\Sigma/M_{0L} = -1.2943, 
\; a_{0L}/M_{0L} = -6.9015, \; R_\Sigma/M_{0L} = 1.1, \; a_{1E} = 0, 
\; M_{1E}/M_{1L} = 0.0407, \; a_{1L}/M_{1L} = -5.5476, \; 
B_\Sigma/M_{1L} = 1.1539, \; \rho_L M_{0L}^2 = 0.0570, \; \rho_E 
M_{0L}^2 = 0.0057.$ 

 \noindent {\it Dust to Dust --- Wormhole Topology} 

We set $M_L = M_L(r)$ and $M_E = M_E(r)$.  Since 
 space-time is no longer empty, not all geodesic coordinate systems 
are equivalent to the comoving one, so a symmetric coordinate coverage 
of the Euclidean region becomes essential to ensure the existence of a 
second transition surface.  From (\ref{etal}) and (\ref{symEuc}) with 
$a_E = 0$ we obtain: 
 \begin{equation} 
    F(r) \equiv \frac{(- f)^{3/2} t_\Sigma}{M_L} = 
     \frac{\sinh\eta_{E\Sigma} + \eta_{E\Sigma}}{\cosh\eta_{E\Sigma}} 
      \equiv D(\eta_{E\Sigma})   \label{FandD}  \end{equation} 
 where the right hand side defines $D(\eta_{E\Sigma})$, and the left 
hand side defines $F(r)$.  For a 
 Kruskal-Szekeres-Schwarzschild type topology (Lorentzian) \cite{ch2}, 
we expect $f$ to take a 
 Novikov-like form, i.e. symmetric, $f(-r) = f(r)$, with $f(0) = -1$ a 
minimum at $r = 0$, and rising monotonically.  To cover the 
asymptotically flat regions at large $R$ requires $f(\pm\infty) = 0$ 
(since $f_E > 0$ doesn't give a bounce).  We expect the mass $M_L$ to 
be minimum at $r = 0$ and rising monotonically to a finite value.  For 
example 
 \begin{equation} f = \frac{-1}{1 + r^2} , \;\;\;\; 
     M_L = \frac{M_{L\,min} + M_{L\,max}r^2}{1 + r^2} 
         \label{no} \end{equation} 
 With these choices and $M_{L\,min} = (1/3) M_{L\,max}$, $F(r)$ and 
$D(\eta_{E\Sigma})$ are plotted in Figs 2a and 2b.  The main features 
of $F(r)$ are dictated by the topology, and are independent of the 
particular choices of $f(r)$ and $M_L(r)$. 

 \begin{center} 
               ================================ \\ 
                  FIGS  2a  AND  2b  GO HERE \\ 
               ================================ 
 \end{center} 

The mapping between $F$ and $D$ is needed to fix the $r$ dependence of 
$\eta_{E\Sigma}(r)$.  We have at our disposal only one constant, 
$t_\Sigma/M_{L\,min}$, which is freely adjustable, so we need to 
select the section of the $D$ graph which includes 0, in order to 
accomodate $f \rightarrow 0$.  Also since $F$ is monotonic, the range 
of $\eta_{E\Sigma}$ cannot extend through the maximum of 
$D(\eta_{E\Sigma})$, $D_{max}$.  Hence we have a restriction on when 
in the Euclidean evolution the transition can occur.  To obtain $f(0) 
= -1$ we need 
 \begin{equation} f = -1 \;\; \Rightarrow \;\; 
\frac{t_\Sigma}{M_{L\,min}} = 
    \frac{\sinh\eta_{E\Sigma} + \eta_{E\Sigma}}{\cosh\eta_{E\Sigma}} 
     \label{ratio}  \end{equation} 
 and consequently the following ranges are allowed: 
 \begin{equation}  0 \leq \frac{t_\Sigma}{M_{L\,min}} \leq D_{max} , 
\;\;\;\; 
       0 \leq \eta_{E\Sigma} \leq \eta_{E\Sigma max} , \;\;\;\; 
       \frac{M_L}{\cosh\eta_{E\Sigma max}} \leq M_E \leq M_L  
  \label{ranges}  \end{equation} 
 where 
 \begin{equation} D_{max} = 1.5434 , \mbox{~~~~~~at~~~~~~} 
\eta_{E\Sigma max} 
        = 1.5434 = D_{max}  \end{equation} 
 Although the range $\eta_{E\Sigma max} \leq \eta_{E\Sigma} \leq 
\infty$ is not obviously precluded in principle, it results in a 
different sign for $d\eta_{E\Sigma}/dr$, which affects $R'_\Sigma$, 
$\rho_{E \Sigma}$ and $\rho_{L \Sigma}$, as well as limiting the range 
of $R$. 

We now find some sample solutions numerically.  Our plotting procedure 
is as follows: \\ 
 \hspace*{5mm}--- Select the Lorentzian mass, $M_L$; \\ 
 \hspace*{5mm}--- Choose a transition time $t_\Sigma$ which complies 
with (\ref{ranges}); \\ 
 \hspace*{5mm}--- Generate values of $M_E$  which span all values 
allowed by (\ref{ranges}), for the given $t_\Sigma$ and $M_L$; \\ 
 \hspace*{5mm}--- For each $M_E$ value calculate: \\ 
 \hspace*{12mm}--- $f$ from (\ref{symEuc}):~~ 
 $f = -[(\sinh\eta_{E\Sigma} + \eta_{E\Sigma})M_E/(t_\Sigma - 
a_E)]^{2/3}$; \\ 
 \hspace*{12mm}--- $r$ from (\ref{no}):~~ 
 $r = \sqrt{(1 + f)/(-f)}$; \\ 
 \hspace*{12mm}--- $R_\Sigma$ from (\ref{con}):~~ 
 $R_\Sigma = (M_L + M_E)/(-f)$; \\ 
 \hspace*{12mm}--- $a_L$ from (\ref{algen}):~~ 
  $a_L = t_\Sigma - (\eta_{L\Sigma} - \sin\eta_{L\Sigma})M_L/(-
f)^{3/2}$; \\ 
 \hspace*{12mm}--- $\rho_{E\Sigma}$ and $\rho_{L\Sigma}$ from  
(\ref{tolden}):~~ 
 $\rho_\Sigma = 2M'/R^2_\Sigma R'_\Sigma$ \\ 
 (The last requires the values of $M_E'$ and $R'$, obtained from the 
derivatives with respect to $r$ of (\ref{symEuc}), (\ref{algen}), 
(\ref{etal}), and (\ref{con})). 

Having found two surfaces where a signature change could occur, i.e. 
$t_\Sigma^{L \rightarrow E} = - t_\Sigma^{E \rightarrow L} > 0$, the 
idea is to excise the future singularity in one Lorentzian region, and 
the past singularity in the other, and join them with the Euclidean 
region. 

The following three models are typical.  They use the forms $M_L = 
(M_{L\,min} + M_{L\,max} r^2)/(1 + r^2)$, $(-1 + f_\infty r^2)/(1 + 
r^2)$ and the values:\\ 
 \hspace*{5mm} (a) \parbox[t]{15cm}{ 
       $t_\Sigma/M_{L\,min} = 1.4096484 = D(1)$ \\ 
       $M_{L\,min}/M_{L\,max} = 0.93$ \\ 
       $f_\infty = -0.93$ 
                                } \\[.1mm] 
 \hspace*{5mm} (b) \parbox[t]{15cm}{ 
       $t_\Sigma/M_{L\,min} = 1.5434 = D_{max}$ \\ 
       $M_{L\,min}/M_{L\,max} = 0.93$ \\ 
       $f_\infty = -0.93$ 
                                } \\[.1mm] 
 \hspace*{5mm} (c) \parbox[t]{15cm}{ 
       $t_\Sigma/M_{L\,min} = 1.4096484 = D(1)$ \\ 
       $M_{L\,min}/M_{L\,max} = 1/3$ \\ 
       $f_\infty = 0$ 
                                } \\[.1mm] 
 Figs. 3 to 6 show $f$, $a_L$, $M_E$, $M_L$, $R_\Sigma$, 
$\rho_{E\Sigma}$ and $\rho_{L\Sigma}$ as functions of $r$, for these 
models. 

 \begin{center} 
           ================================================= \\ 
           FIGS  3, 4, 5a, 5b, 5c, 6a, 6b,  AND  6c  GO HERE \\ 
           ================================================= 
 \end{center} 

As expected, the areal radius has a minimum at $r = 0$ and does not go 
singular, and both the Lorentzian and Euclidean `densities' are well 
behaved at the transition.  

     Model (a) has $\Sigma$ entirely inside the external event horizon 
$R = 2 M_{L\,max}$, as shown in Fig.\ 5a, and it was found that there 
must be very little variation in $M_L(r)$ and $f(r)$ (Fig.\ 3) to 
achieve this.  The densities on each side of $\Sigma$ are also only 
mildly varying, as shown in Fig.\ 6a.  It could be thought of as a 
perturbation of a 
 Kantowski-Sachs model, allowing the particle world lines to emerge 
beyond $R = 2M_L$ briefly, before recollapsing back inside and 
encountering the signature change.  Since this is true even for $r 
\rightarrow \infty$, the particles do not fill the spacetime, and the 
model may be completed by matching to a vacuum exterior.  This makes a 
very satisfactory model of signature change in the black hole 
topology. 

     Model (b) differs only in having the largest possible value of 
$t_\Sigma/M_{L\,min}$.  This results in $a_L$, $M_E$, $M_L$, 
$R_\Sigma$, $\rho_{E\Sigma}$ and $\rho_{L\Sigma}$ all having 
 non-zero gradient at $r = 0$, meaning these quantities are 
discontinuous through the origin
 --- see Figs.\ 5b and 6b.  A feature of this model is that the bang 
time $a_L(r)$ has a maximum away from $r = 0$ (Fig.\ 4), indicating 
the particle world lines self intersect somewhere in the time 
evolution of the Lorentzian part of the model \cite{shcr,ch2}.  This 
`shell crossing' is now a serious deficiency of the model, involving 
densities that diverge and go negative.  Otherwise it is very similar 
to (a). 

     Model (c) is complete since $r \rightarrow \infty$ covers the 
asymptotic regions of the model, and $f \rightarrow 0$ means it is 
asymptotically flat.  However the signature change surface passes out 
of $R = 2 M_{L\,max}$ and extends to $R = \infty$ (Fig.\ 5c), which is 
not ideal, despite the nice density profile in Fig.\ 6c.

 \noindent {\it Summary} 

Within the 
 Lema\^{\i}tre-Tolman metric form, the wormhole topology is possible 
in both Lorentzian and Euclidean regions with and without matter 
(dust) present.  The Euclidean region bounces provided the mass 
function is positive (as defined in (\ref{tolvel})), so negative mass 
models were not considered.  Only constant $t$ transition surfaces, 
which are orthogonal to the fluid flow, were considered.  

Vacuum to Vacuum signature transitions are equivalent to the 
uninteresting constant $T$ Schwarzschild case.  

Signature transitions {\em are} possible inside the horizon if the 
density is non zero on both sides of the transition.  Time symmetric 
Euclidean regions permit a second transition following the bounce, 
emerging into an expanding spacetime behind the past singularity.  
This is true both for the constant $R$ transitions in uniform density 
 Kantowski-Sachs type models, and the more general inhomogeneous case.  
The general case is particularly satisfactory as it models a finite 
cloud of dust. 

 \noindent {\large \bf 5.~ Conclusion} 

  We have succeeded in demonstrating the possibility that a change in 
the signature of 
 space-time may occur in the late stages of black hole collapse, 
resulting in a Euclidean region which bounces and 
 re-expands, passing through a second signature change to a new 
expanding Lorentzian 
 space-time.  The classical singularity at $R=0$ is thus avoided.  
Such transition surfaces necessarily have 
 non-zero extrinsic curvature. 

  The model of signature change employed here is strictly classical.  
Quantum cosmological questions, for example the relative probability 
of different sorts of transitions, have not been considered.  We have 
based our notion of manifold continuity on the fulfillment of the 
 Darmois type matching conditions, since they are invariant to the 
coordinates used, and no modifications are necessary to adapt them to 
surfaces of signature change.  As discussed in \cite{cont}, surface 
effects appear in the conservation laws, even when stronger conditions 
than Darmois' are imposed. 

  Based on this approach, we have shown that signature transitions are 
possible in a spherically symmetric Lorentzian 
 space-time, in both the Schwarzschild and 
 Lema\^{\i}tre-Tolman metric representations, though the ensuing 
Euclidean region might not be empty.  Once the Israel identities are 
adapted to signature change, continuous `density' is no longer 
required. 

  Within the Schwarzschild metric form, such a transition was possible 
on a constant $T$ slice, but this can only span the outer region $R 
\geq 2M$. Conversely the constant $R$ surface can be entirely inside 
the horizon, but does not lead to a bouncing Euclidean region.  Thus 
these models are not satisfactory.  

  A study of the geodesics in each region showed that the two 
Euclidean regions, $R \geq 2M$ and $R \leq 2M$ were in fact complete 
manifolds.  It was found necessary to match geodesic 
 4-momenta, $P^\mu$, at the signature change, in order that all 
geodesics could be continued.  This naturally means $P_\mu$ and $m^2 = 
|P^\mu P_\mu|$ are discontinuous.  This is consistent with the fact 
that the density can jump at a signature change. 

  These results were generalised using constant $t$ transitions in the 
 Lema\^{\i}tre-Tolman metric form.  With suitable choices of the 
function $f(r)$, this metric can reproduce the 
 Kruskal-Szekeres topology of two sheets joined by a wormhole, but 
with non zero density.  It also has a 
 Kantowski-Sachs limit.  It was found possible to have a signature 
change surface completely hidden inside the horizon $R = 2 M$ in the 
Lorenzian region, provided there was 
 non-zero density in both the Lorentzian and Euclidean regions.  In 
the Lorentzian region, the matter is of finite extent, and may be 
surrounded by vacuum.  It was also possible for the Euclidean region 
to be 
 time-symmetric, so that after the bounce, the matter expands through 
a second signature change into another Lorentzian region 
 --- a new universe. This makes a very satisfactory model of collapse, 
bounce and 
 re-expansion of a mass concentration.   Within the 
 Lema\^{\i}tre-Tolman form, a constant $t$ signature change surface 
cannot be arbitrarily close to the Lorentzian singularity $R = 0$.  
One might expect such transitions to occur only a Planck time before 
$R = 0$, which would require us to consider a different equation of 
state in the Euclidean region.  This may well relax the limits on 
$\eta_\Sigma$ and $f$ that were found. 

It was found possible to hide the entire signature change surface 
inside the Lorentzian horizon $R = 2M_L$, if the model is 
 non-vacuum in the central regions, with a vacuum exterior.  The 
matter is collapsing from not far outside the horizon, as may be 
expected for a collapsing compact object.  The limit on $f$ for a 
completely hidden surface implies (1) that all the infalling matter 
must be in a finite cloud, moving on tightly bound paths ($R_{max} 
\leq 4M_L$), surrounded by vacuum, and (2) that the black hole 
topology is required.  This provides a classical bounce model of the 
kind we sought.  A quantum cosmological analogue could be of interest 
in the context of Smolin's `natural history' of universes proposal.  
A treatment similar to that of Kerner and Martin 
\cite{KernMart,Martin} could permit the 
 non-zero extrinsic curvature that is required.

 \noindent {\large \bf Acknowledgement} 

     AS thanks the FRD for a Bursary.  CH thanks the FRD for a Grant, 
and Robin Tucker and the Physics Department at Lancaster University 
for hospitality.

 \newpage 

 \noindent {\large \bf Tables} 

 \begin{table}[h] 
 \begin{center} 
 \begin{tabular}{|c|c|c|} \hline 
 Continuation & \multicolumn{2}{c|}{ Signature Change Surface }\\ 
\cline{2-3} 
 Condition    & Const. $R$, $R < 2M$ & Const. $T$, $R \geq 2M$ \\ 
\hline 
   (i) $u^R_E = u^R_L$ & $q_E = q_L$,~~~$h_E^2 = - h_L^2 = 0$ 
          & $q_E = q_L$,~~~$h_E^2 + h_L^2 = 2(1 - 2M/R)$  \\ \hline 
  (ii) $u^T_E = u^T_L$ & $h_E = h_L$ & $h_E = h_L$ \\ \hline 
 (iii) $P^\mu_E = P^\mu_L$ & 
    $2 h_E^2 h_L^2 = (h_L^2 - h_E^2) (2M/R - 1)$, & 
    $2 h_E^2 h_L^2 = (h_L^2 + h_E^2) (1 - 2M/R)$, \\ 
      & $q_E = q_L$,~~~$m_E h_E = m_L h_L$ 
      & $q_E = q_L$,~~~$m_E h_E = m_L h_L$ \\ \hline 
 \end{tabular} 
 \caption{Conditions resulting from the three continuation conditions 
at two types of signature change that retain a Schwarzschild metric 
form. \label{GeodCont} } 
 \end{center} 
 \end{table} 

 \begin{table}[h] 
 \begin{center} 
 \begin{tabular}{|l|c|c|} \hline 
 Transition surface & Const. $R$, $R < 2M$ & Const. $T$, $R \geq 2M$ 
   \\ \hline 
 Allowed $h_L$ & $h_L^2 \geq 0 \geq (1 - 2M/R)$ 
                         & $h_L^2 \geq (1 - 2M/R)$ \\ 
   \hline 
 Allowed $h_E$ & $h_E^2 \leq (2M/R - 1)$ & $h_E^2 \leq (1 - 2M/R)$ \\ 
   \hline 
 \end{tabular} 
 \caption{Summary of allowed ranges of geodesic energy parameters on 
either side of the two signature change surfaces.  \label{hRanges} } 
 \end{center} 
 \end{table} 

 \noindent {\large \bf Figure Captions} 


 {\bf Fig. 1.~~} Diagram illustrating properties of combined Euclidean 
and Lorentzian geodesic paths.  The curves give values of various 
parameters, as functions of $h_L$ for geodesics arriving at a constant 
$T$ signature change surface at the point $R_\Sigma = 4M$(those curves 
starting at $h_L = 1/\sqrt{2}$), and at constant $R = 4M/5$ surface 
(those curves starting at $h_L = 0$).  Vertical slices through the 
graph give the values of $h_L$, $h_E$, $m_L/m_E$, $R^{max}_L$ (where 
it exists) and $R^{min}_E$ or $R^{max}_E$ for individual geodesics. 

 {\bf Fig. 2a.~~} The function $F(r)$ vs. $r$. 

 {\bf Fig. 2b.~~} The function $D(\eta)$ vs. $\eta$. 

 {\bf Fig. 3.~~} $f(r)$ vs. $r$ for the three Lema\^{\i}tre-Tolman 
signature change models, (a), (b) and (c). 

 {\bf Fig. 4.~~} $a_L(r)$ vs $r$, for models (a), (b) and (c).  Note 
that (b) has its maximum away from $r=0$, whereas (a) and (c) has it 
at $r = 0$. 

 {\bf Fig. 5a.~~} The run of $R_\Sigma(r)$, $M_L(r)$ and $M_E(r)$ for 
model (a). 

 {\bf Fig. 5b.~~} $R_\Sigma(r)$, $M_L(r)$ and $M_E(r)$ for model (b).  
Note that none of these quantities are smooth through the origin $r = 
0$.  

 {\bf Fig. 5c.~~} $R_\Sigma(r)$, $M_L(r)$ and $M_E(r)$ for model (c). 

 {\bf Fig. 6a.~~} The run of $\rho_L$ and $\rho_E$ for model (a). 

 {\bf Fig. 6b.~~} $\rho_L$ and $\rho_E$ for model (b).  Note that 
neither quantity is smooth through the origin $r = 0$. 

 {\bf Fig. 6c.~~} $\rho_L$ and $\rho_E$ for model (c). 

 \end{document}